\documentclass[useAMS,usenatbib,usegraphicx]{mn2e}

\title[Multi-wavelength observations of GRB~090102]{Testing GRB models with the afterglow of GRB~090102}

\author[B. Gendre et al.] {
B. Gendre$^{1}$\thanks{Present address : ASDC, Via Galileo Galilei, 00044 Frascati, Italy. E-mail:bruce.gendre@asdc.asi.it. Based on observations obtained with TAROT, REM, GROND.},
A. Klotz$^{2,3}$,
E. Palazzi$^{4}$,
T. Kr\"uhler$^{5,6}$,
S. Covino$^{7}$,
P. Afonso$^{5}$,
\newauthor
L.A. Antonelli$^{8}$,
J.L. Atteia$^{9}$,
P. D'Avanzo$^{7}$,
M. Bo\"er$^{2}$,
J. Greiner$^{5}$,
and S. Klose$^{10}$
\\
$^{1}$Laboratoire d'Astrophysique de Marseille/Université de Provence/CNRS, 38 rue Joliot curie, 13388 Marseille CEDEX 13, France\\
$^{2}$Observatoire de Haute-Provence, F--04870 Saint Michel l'Observatoire, France\\
$^{3}$CESR, Observatoire Midi-Pyr\'en\'ees, CNRS, Universit\'e de Toulouse, BP 4346, F--31028 - Toulouse Cedex 04, France\\
$^{4}$IASF-Bologna/INAF, via Gobetti 101, 40129, Bologna, Italy\\
$^{5}$Max-Planck-Institut f\"ur extraterrestrische Physik, Giessenbachstrasse 1, 85748 Garching, Germany\\
$^{6}$Universe Cluster, Technische Universit\"at M\"unchen, Boltzmannstrasse 2, 85748 Garching, Germany\\
$^{7}$Osservatorio Astronomico di Brera/INAF, via Bianchi 46, 23807 Merate (LC), Italy\\
$^{8}$Osservatorio Astronomico di Roma/INAF, via Frascati 33, 00040 Monteporzio Catone (RM), Italy\\
$^{9}$LATT, Observatoire Midi-Pyr\'en\'ees, CNRS, Universit\'e de Toulouse, 14 Avenue E. Belin, F--31400 - Toulouse, France\\
$^{10}$Th\"uringer Landessternwarte Tautenburg, Sternwarte 5, 07778 Tautenburg, Germany
}

\begin{document}

\date{Accepted 2010 Feb 28. Received 2010 Feb 18; in original form 2009 Sept 07}

\pagerange{\pageref{firstpage}--\pageref{lastpage}} \pubyear{2009}

\maketitle

\label{firstpage}

\begin{abstract}

We present the observations of the afterglow of gamma-ray burst GRB~090102. Optical data taken by the TAROT, REM, GROND, together with publicly available data from Palomar, IAC and NOT telescopes, and X-ray data taken by the XRT instrument on board the \emph{Swift} spacecraft were used. This event features an unusual light curve. In X-rays, it presents a constant decrease with no hint of temporal break from 0.005 to 6 days after the burst. In the optical, the light curve presents a flattening after 1 ks. Before this break, the optical light curve is steeper than that of the X-ray. In the optical, no further break is observed up to 10 days after the burst. We failed to explain these observations in light of the standard fireball model. Several other models, including the cannonball model were investigated. The explanation of the broad band data by any model requires some fine tuning when taking into account both optical and X-ray bands.
\end{abstract}


\section{Introduction}

Gamma-Ray Bursts \citep[GRBs, see e.g.][for a review]{piran06} were discovered in the late 1960's \citep{kle73}. For about three decades, their exact nature remained elusive. The effort to provide a fast re-pointing of the BeppoSAX satellite allowed for the first detection of their afterglows at all wavelengths \citep[e.g.][]{cos97,van97}. Since then, the specification of each generation of instrumentation devoted to GRB studies included the need to point to the GRB afterglows quickly. Now, with Swift, we have reached the point where we can collect data from X-rays to optical within seconds after the start of the event, thanks to fast moving telescopes triggered by GCN notices \citep{Barthelmy1998}. This run toward fast re-pointing has allowed for prompt optical emission observations by small autonomous telescopes for several GRBs \citep[e.g.][]{klo09c}.

Since the mid of the 1990's, a theoretical canonical model has emerged in order to explain the GRB phenomenon: the fireball model \citep{ree92, mes97, pan98}. This model is based on blast waves emitted by a progenitor during black hole formation, that interacts within a jet and/or with the surrounding medium to produce all observed emissions. These emissions are not thermal, and are proposed to be synchrotron emission from the electrons accelerated within the fireball. This model has the advantages that it is quite easy to apply and reproduces, reasonably well, the observations done by BeppoSAX and other telescopes \citep[e.g.][]{gen06}. The situation has, however, radically changed with the launch of the {\it Swift} satellite \citep{ghe04}. {\it Swift} has discovered X-ray flares \citep{obr05}, an early flattening of the X-ray emission \citep{nou04}, and several chromatic breaks within the afterglow light curves \citep[e.g.][]{rac08}, that were unexpected within the initial standard model. This model has been modified in order to explain all these new features, including late energy injection \citep[e.g.][]{ghi07}, multi-core or patchy jets \citep[e.g.][and references therein]{pen05}, and multi-component afterglow \citep[e.g.][]{wil06}. However, these modifications to the standard model were made only in order to explain a given feature, and sometimes lack a multi-wavelength consistency check. In previous years, several studies on peculiar events have led the realization that all available data could not be explained due to limitations within the standard model \citep[e.g.][]{klo08}.

In this article, we present observations of GRB 090102 at optical and X-ray wavelengths, taken over the 10 days following the burst. While the optical or X-ray data alone could be explained within the framework of the standard model, when combining these different datasets, the standard model alone cannot explain the observations of this burst, indicating that other models need to be tested. The paper is organized as follows: we present GRB 090102 in Sec. \ref{sec_burst} with the data reduction and analysis in Sec. \ref{sec_data}; we fit several models to these data in Sec. \ref{sec_model}; finally, we discuss our findings in Sec. \ref{sec_discu} before concluding. Errors on spectral and temporal indices (and the related closure relation values) are quoted at the 90\% confidence level throughout. Reported errors on data points are $1\sigma$ errors.

\section{GRB~090102}
\label{sec_burst}

GRB~090102 triggered {\it Swift}-BAT \citep[trigger 338895,][]{man09a} on Jan. 2009, 2$^{nd}$ at 02:55:45 UT  (hereafter, $T_0$). It was also recorded by Konus Wind \citep{gol09}, and by INTEGRAL/SPI-ACS \citep{man09b}. The duration of this event, as observed by {\it Swift} was T$_{90}$ = $27.0 \pm 2.2$ s. Its light curve presents four peaks from $T_0-14$ s to $T_0+15$~s \citep{sak09}. Its time averaged spectrum can be represented either using an exponential cutoff power law model or a Band model \citep{band93}, with parameters $\alpha = -0.86 \pm 0.14$, $\beta \le -2.73$, and $E_p = 451_{-58}^{+73}$ keV ($\chi^2_\nu = 1.075$, 64 d.o.f.) and with a fluence of $3.09_{-0.25}^{+0.29} \times 10^{-5}$ erg cm$^{-2}$ \citep[20~keV -- 2 MeV,][]{gol09}. Using the measured redshift \citep[$z$=1.547,][]{pos09a}, this gives $E_{\rm iso} \sim 5.75 \times 10^{53}$ erg (1~keV -- 10 MeV, rest frame) and a rest frame $E_p$ of $1149_{-148}^{+186}$ keV, fully compatible with the so-called Amati relation \citep{ama09, ama02}.

\begin{figure}
\centering
  \includegraphics[width=8.6cm]{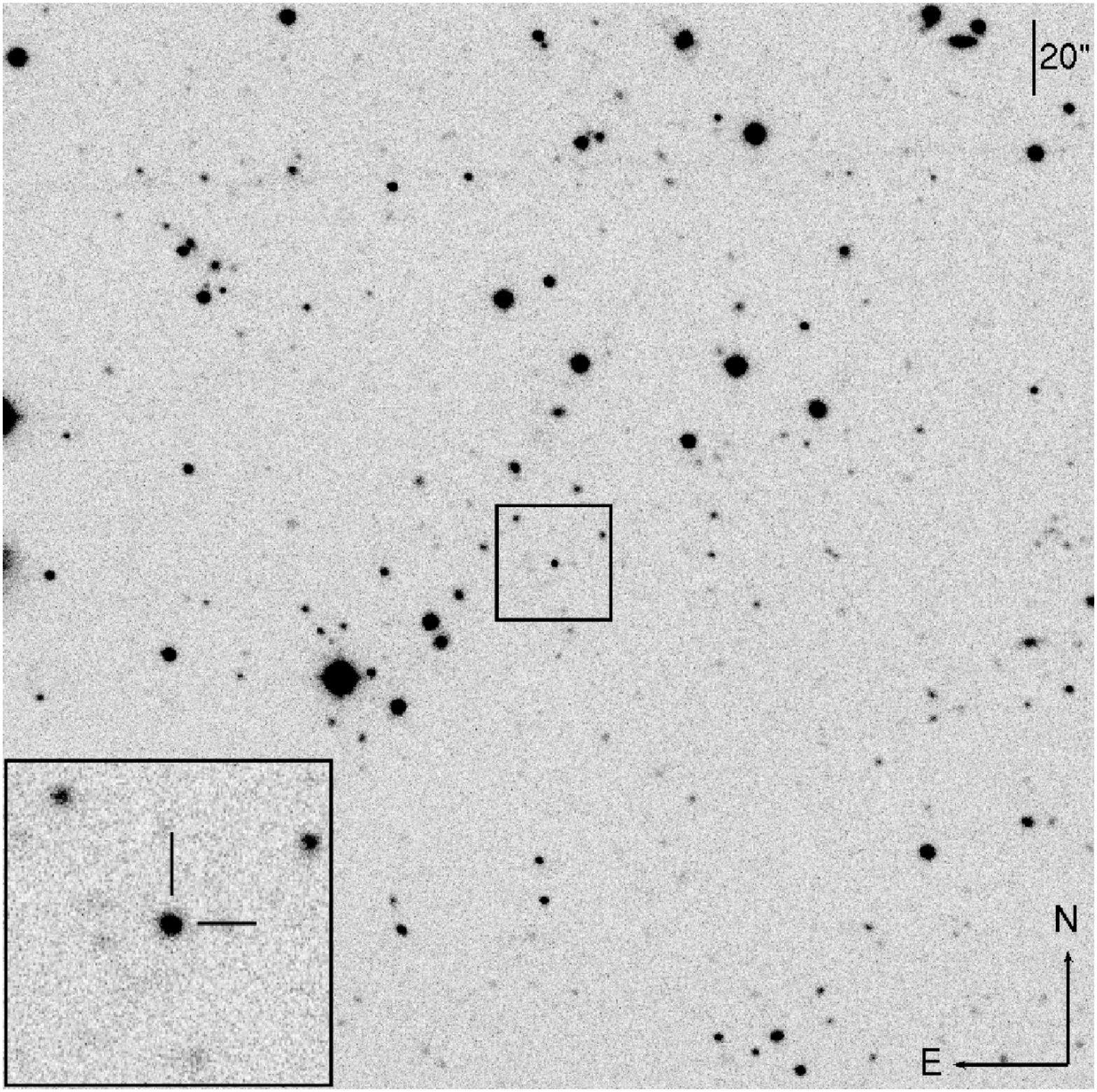}
  \caption{Image of the field of view of GRB 090102 from GROND, taken in the i' band 11 ks after the burst. The GRB afterglow is indicated by two ticks.\label{image_champ}}
\end{figure}

Due to an observational constraint (Earth limb too close), \emph{Swift} slewed to the position of the afterglow with a delay, and XRT and UVOT observations began only at $T_0+395$ s, observing a bright fading source in both the XRT and the UVOT fields of view \citep{man09a}. No optical observations were available during the prompt phase. The earliest ground search of the optical afterglow was performed by the Telescope a Action Rapide pour les Objets Transitoires (TAROT), starting at $T_0$+40.8 s \citep{klo09a}. The Rapid Eye Mount (REM) robotic telescope also observed the field starting from $T_0$+53 s \citep[][note however that the start time reported in the GCN is not correct]{cov09}. A new source was detected at RA=08$^h$32$^m$58.52$^s$, Dec=+33$^\circ$06\arcmin51.1\arcsec (J2000.0) with an error of 0.3\arcsec (see Fig. \ref{image_champ}). 
This source was then confirmed to be the optical afterglow of GRB 090102. Spectroscopic observations were carried by the Nordic Optical Telescope (NOT) \citep{pos09a}, the Palomar 200" Hale telescope \citep{cen09}, and by the Hobby-Eberly Telescope \citep{cuc09}.

Late afterglow observations were performed by the Gamma-Ray burst Optical/Near-Infrared Detector (GROND) 2.50 h after the burst \citep{afo09}, by the Instituto de Astrof\'isica de Canarias 80 cm telescope (IAC80) 19.2 h after the burst \citep{pos09b}, and by the Palomar 60-inch telescope 50 min after the burst and later on \citep{cen09}. The observations continued in the following days using the NOT \citep{mel09} and the HST \citep{lev09}, leading to the detection of the host galaxy.

A radio follow-up was also carried out using the VLA and the Westerbork Synthesis Radio Telescope (WSRT) at the 8.46 GHz \citep{cha09} and the 4.9 GHz \citep{hor09} wavelengths respectively. No afterglow was detected, leading to upper limits. Finally, the MAGIC Cerenkov Telescope also pointed to the position of the afterglow, without any detection \citep{gau09}.

\section{Data reduction and analysis}
\label{sec_data}

\subsection{X-ray data}

\begin{figure*}
\centering
  \includegraphics[width=16cm]{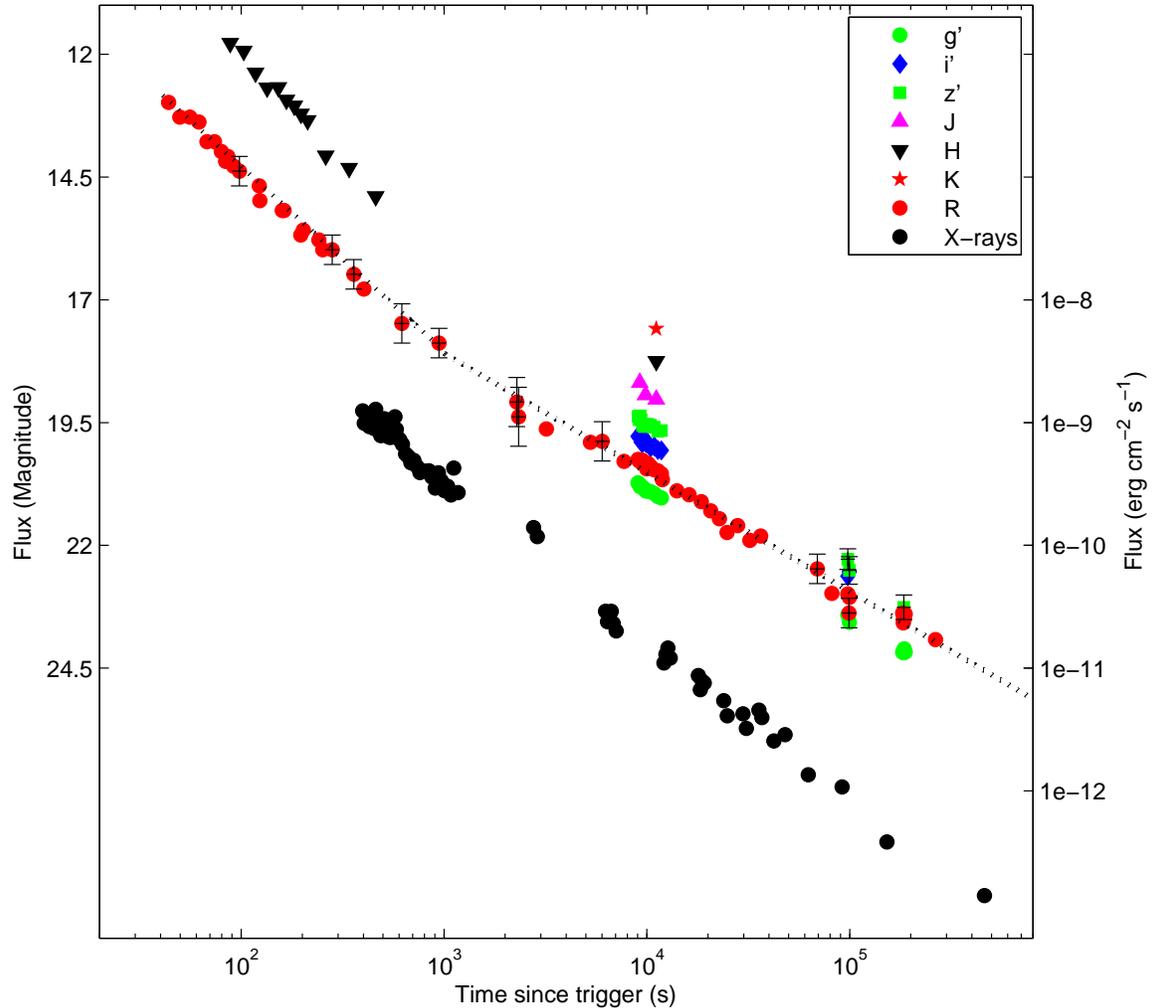}
  \caption{\label{lc_all} Light curve of the afterglow of GRB 090102 in X-ray, g', R, i', z', J, H, and K bands from Tables \ref{table_data_R} and \ref{table_data}. For clarity, upper limits have been omitted. When not plotted, the error bars have the size of the symbols or smaller. These light curves have been corrected for the galactic extinction/absorption due to the Milky Way. The X-ray light curve is extracted in the 0.5 to 10.0 keV band. The dotted line is the best fit decay law (see text for details). The left axis applies to optical/infrared data, the right axis to X-ray data.}
\end{figure*}

The data were processed using the ftools version 6.6.2. To extract light curves and spectra, a rectangular box and circular extraction region were used for window timing and photon counting modes respectively, with an extraction region of 50 pixels in both cases. XRT observations began 395 seconds after $T_0$ in window timing mode. The instrument switched to the photon counting mode \citep[see][]{hil05} at $T_0 + 669.1$ s, while the count rate was high enough to cause pile up. As a matter of consequence, the observation between $T_0 + 669.1$ s and $\sim T_0 + 3000$ s suffers from severe to moderate pile up. This area was discarded in the following spectral analysis, and corrected within the light curve using the prescriptions of \citet{vau05}. 

All spectra were rebinned in order to include 20 net counts within each bin, and fit using the $\chi^2$ statistic. The spectral model was composed of a power law continuum absorbed by our own galaxy \citep[$N_H = 4.1 \times 10^{20}$ cm$^{-2}$,][]{dic90} and by the host galaxy at z = 1.547. Extracting spectra at several epoch, we did not find any evidence for spectral variation during the whole follow-up of the afterglow. The spectrum is well modeled ($\chi^2_\nu = 0.92$, 102 d.o.f.) by a single power law with spectral index $\beta_X = 0.83 \pm 0.09$ absorbed in our own galaxy and by extragalactic absorbers ($N_H = 8.8_{-2.5}^{+2.7} \times 10^{21}$  cm$^{-2}$ when located within the host galaxy).

The light curve was extracted within the 0.5-10.0 keV band and rebinned in order to obtain at least 25 counts per bin. All decay index indicated below are derived from fits using the $\chi^2$ statistic. We did not observe strong flares (see Fig. \ref{lc_all}). A single moderate flare can be seen in the unbinned light curve during orbit 2 (corresponding to the period $\sim 2000-4000$s after the trigger), which has been excluded from the temporal analysis. The complete light curve can be adequately fit ($\chi^2_\nu = 1.18$, 65 d.o.f.) using a single power law with a decay index $\alpha_X = 1.34 \pm 0.02$. A broken power law can also represent the data ($\chi^2_\nu = 0.99$, 63 d.o.f.) with $\alpha_{1,X} = 1.29 \pm 0.03$, $\alpha_{2,X} = 1.48 \pm 0.10$, and a break time $t_b = 18700^{+14500}_{-8000}$ s. With the observed steepening, $\Delta\alpha = 0.19 \pm 0.11$, this break could be interpreted as the cooling frequency passing through the observation band. However, this should be associated with a spectral break not supported by the data. An F-test check on the break existence gives a probability of 0.26, i.e. a value not conclusive. One may note however that in this special case (large error on $t_b$ that make it compatible with 0 within $3\sigma$), the probability derived from the F-test should not be considered as valid \citep[see][]{pro04}. As a matter of consequence, in the following we will consider both hypotheses (single power law or broken power law), and report early and late X-ray data as data taken before and after the temporal break respectively.


\subsection{Optical data}

At the position of the afterglow, the Galactic extinction is $E(B-V) = 0.047$ according to \citet{sch98}. We corrected all magnitudes using this estimation. Assuming $R_v$=3.1, this gives in particular $A_{\rm R}$=0.14 \citep{pei92}. The specific data processing relative to all instruments is indicated in the following subsections. We completed this sample by using reported observations by \citet{cen09, pos09b, mel09}. This extended sample of data taken in the R band is listed in Table \ref{table_data_R}.

\begin{table*}
\caption{Table of data taken in the R band, corrected for galactic extinction. GROND data in r' band are transformed to Vega R system. Errors are given at $1\sigma$ level. Data from Palomar, IAC80 and NOT are taken from \citet{cen09, pos09b, mel09} respectively.}
\label{table_data_R}
\centering
\begin{tabular}{ccccc|ccccc}
\hline
Mid time  & Exposure &  Magnitude & Instrument    & \phantom{tex}& \phantom{tex}& Mid time  & Exposure &  Magnitude & Instrument\\
          & duration &            &               & & &           & duration &            & \\
  (s)     & (s)      &            &               & & &  (s)      & (s)      &            & \\
\hline
  43.8    &   6.0    &  13.0  $\pm$  0.2 & TAROT  & & & 9266      & 66       &  20.43 $\pm$ 0.03 & GROND \\
  49.8    &   6.0    &  13.3  $\pm$  0.2 & TAROT  & & & 9373      & 66       &  20.48 $\pm$ 0.06 & GROND \\
  55.8    &   6.0    &  13.3  $\pm$  0.2 & TAROT  & & & 9514      & 115      &  20.44 $\pm$ 0.04 & GROND \\
  61.8    &   6.0    &  13.4  $\pm$  0.2 & TAROT  & & & 9711      & 115      &  20.51 $\pm$ 0.02 & GROND \\
  67.8    &   6.0    &  13.8  $\pm$  0.2 & TAROT  & & & 9903      & 115      &  20.52 $\pm$ 0.02 & GROND \\
  73.8    &   6.0    &  13.8  $\pm$  0.2 & TAROT  & & & 9973.9    & 120      &  20.44 $\pm$ 0.07 & Palomar \\
  79.8    &   6.0    &  14.0  $\pm$  0.2 & TAROT  & & & 10091     & 115      &  20.50 $\pm$ 0.03 & GROND \\
83.8      & 30       &  14.18 $\pm$ 0.10 & ROSS   & & & 10423     & 375      &  20.56 $\pm$ 0.02 & GROND \\
  85.8    &   6.0    &  14.1  $\pm$  0.2 & TAROT  & & & 10868     & 375      &  20.63 $\pm$ 0.01 & GROND \\
  91.8    &   6.0    &  14.3  $\pm$  0.2 & TAROT  & & & 11320     & 375      &  20.65 $\pm$ 0.01 & GROND \\
  97.8    &   6.0    &  14.4  $\pm$  0.3 & TAROT  & & & 11780     & 375      &  20.71 $\pm$ 0.02 & GROND \\
 122.7    &  29.4    &  14.7  $\pm$  0.1 & TAROT  & & & 11911.5   & 120      &  20.66 $\pm$ 0.08 & Palomar \\
123.6     & 30       &  14.98 $\pm$ 0.20 & ROSS   & & & 14068.6   & 120      &  20.89 $\pm$ 0.09 & Palomar \\
 159.3    &  30.6    &  15.2  $\pm$  0.2 & TAROT  & & & 16144.7   & 120      &  20.97 $\pm$ 0.09 & Palomar \\
162.4     & 30       &  15.18 $\pm$ 0.20 & ROSS   & & & 18522.6   & 600      &  21.11 $\pm$ 0.06 & Palomar \\
 196.5    &  30.6    &  15.7  $\pm$  0.2 & TAROT  & & & 20631.5   & 600      &  21.30 $\pm$ 0.06 & Palomar \\
202.2     & 30       &  15.58 $\pm$ 0.20 & ROSS   & & & 22756.7   & 600      &  21.46 $\pm$ 0.07 & Palomar \\
241.1     & 30       &  15.78 $\pm$ 0.20 & ROSS   & & & 24832.6   & 600      &  21.74 $\pm$ 0.08 & Palomar \\
 252.3    &  67.8    &  16.0  $\pm$  0.2 & TAROT  & & & 27994.7   & 1200     &  21.60 $\pm$ 0.05 & Palomar \\
280.8     & 30       &  15.98 $\pm$ 0.30 & ROSS   & & & 32188.5   & 1200     &  21.90 $\pm$ 0.06 & Palomar \\
358.6     & 90       &  16.48 $\pm$ 0.30 & ROSS   & & & 36411.3   & 1200     &  21.81 $\pm$ 0.11 & Palomar \\
 402.3    & 185.4    &  16.8  $\pm$  0.2 & TAROT  & & & 69120.0   & 1800     &  22.48 $\pm$ 0.30 & IAC80 \\
619.5     & 330      &  17.48 $\pm$ 0.40 & ROSS   & & & 81566     & ---      &  22.98 $\pm$ 0.20 & NOT \\
 943.8    & 685.2    &  17.9  $\pm$  0.3 & TAROT  & & & 97805     & 1500     &  23.16 $\pm$ 0.07 & GROND \\
2283.0    &2173.2    &  19.1  $\pm$  0.5 & TAROT  & & & 99200     & ---      &  23.38 $\pm$ 0.30 & IAC80 \\
2328.5    & 1380     &  19.38 $\pm$ 0.60 & ROSS   & & & 99600     & 1500     &  23.23 $\pm$ 0.10 & GROND \\
3190.2    & 120      &  19.63 $\pm$ 0.13 & Palomar& & & 182130    & 1500     &  23.55 $\pm$ 0.12 & GROND \\
5266.2    & 120      &  19.90 $\pm$ 0.06 & Palomar& & & 183448    & ---      &  23.58 $\pm$ 0.20 & NOT \\
6014.4    &5275.2    &  19.9  $\pm$  0.4 & TAROT  & & & 183945    & 1500     &  23.67 $\pm$ 0.11 & GROND \\
7704.3    & 120      &  20.29 $\pm$ 0.07 & Palomar& & & 185758    & 1500     &  23.67 $\pm$ 0.11 & GROND \\
9042      & 66       &  20.42 $\pm$ 0.04 & GROND  & & & 187575    & 1500     &  23.57 $\pm$ 0.09 & GROND \\
9158      & 66       &  20.45 $\pm$ 0.03 & GROND  & & & 264553    & ---      &  23.92 $\pm$ 0.09 & NOT \\
\hline
\end{tabular}
\end{table*}

\subsubsection{TAROT data}

We used data from TAROT Calern \citep{klo09d} that started an exposure of the field of GRB~090102 at T$_0+$40.8~s (duration 60 s) with the tracking speed adapted to obtain a small ten pixel trail. This technique was used in order to obtain continuous temporal information during the exposure \citep[e.g.][]{klo06}. The spatial sampling was 3.29\arcsec pix$^{-1}$ and the FWHM of stars (in the perpendicular direction of the trail) was 2.05 pixels. Only the first exposure was performed with this technique. Successive images were tracked on the diurnal motion using exposure times increasing from 30 s to 180 s. Images are not filtered.

We used the star USNO-B1 1231-0206902 as a constant template for calibration during the trailed exposure. Contrary to the claim of \citet{klo09a}, TAROT did not observe a rise during the first part of the trailed image. Indeed, the profile is convolved by the Point Spread Function (PSF) that induces a slope at the start and at the end of the trail. The rapid analysis done by \citet{klo09a} confused this slope with a rise at the start of the GRB afterglow trail. This has been corrected for by using the flux of the template (constant) star.

In order to precisely cross-calibrate the observations performed with REM and TAROT, we used several USNO stars as relative standards.
All flux measurements were done using standard aperture photometry with the AuDela software\footnote{http://www.audela.org/}. USNO-B1 1231-0206902 has been used to build the PSF in TAROT images in order to fit the flux of the GRB. TAROT observations are not filtered. We thus converted all TAROT instrumental magnitudes to standard R magnitudes, using the following formula: R$ = C - 2.5 \times log (F) + ({\rm V-R}) \times \alpha_r$. In this equation, $\alpha_r = 0.4$ is determined using calibration stars with various (V-R) values located within an open cluster. $C$ is calculated for each image from the PSF flux, the R and (V-R) measurements of the reference stars.

\subsubsection{REM data}
Early time optical and NIR data were collected using the 60 cm robotic telescope REM \citep{zer01,cov04} located at the ESO La Silla observatory (Chile). The telescope simultaneously feeds, by means of a dichroic, the two focal instruments: REMIR \citep{con04} a NIR camera, operating in the range 1.0-2.3 $\mu$m (z', J, H and K), and ROSS \citep[REM Optical Slitless Spectrograph,][]{tos04} an optical imager with spectroscopic (slitless) and photometric capabilities (V, R, I). Both cameras have a field of view of 10 $\times$ 10 \arcmin. 

REM reacted promptly and began observing 103 s after the GRB trigger time, following the event for $\sim $1~hour (see Table \ref{table_data_R}). For the first 500~s REMIR observations were performed only in the H filter with increasing exposure times, then all the NIR filters were used in sequence. In the optical range only R band images were taken, for a total of 38 consecutive images. The transient was only detected up to 500 s both in the R and H bands, after which its luminosity falls below the instrument detection limits for all filters.

Each single NIR observation with REMIR was performed with a dithering sequence of five images shifted by a few arcseconds. These images are automatically elaborated using the proprietary routine AQuA \citep{tes04}. The script aligns the images and co-adds all the frames to obtain one average image for each sequence.

Each R image was reduced using standard procedures. A combination of the ESO-MIDAS\footnote{http://www.eso.org/projects/esomidas/} DAOPHOT task \citep{ste88} and Sextractor\footnote{http://astroa.physics.metu.edu.tr/MANUALS/sextractor/} package \citep{ber96} was used to perform standard aperture photometry. The photometric calibration for the NIR was accomplished by applying average magnitude shifts computed using the 2MASS\footnote{http://www.ipac.caltech.edu/2mass/} catalogue.

The optical data were calibrated using instrumental zero points and checked with observations of standard stars in the SA92 Landolt field \citep{lan92}. Results are displayed in Tables \ref{table_data_R} and \ref{table_data}.

\subsubsection{GROND data}
The Gamma-Ray Burst Optical/NearInfrared Detector \citep[GROND,][]{gre08} mounted at the 2.2~m ESO/MPI telescope at La Silla observatory imaged the field of GRB 090102 simultaneously in g'r'i'z'JHK starting 9.01~ks after T$_{\rm 0}$ under clear sky conditions \citep{afo09}. At that point the location of the burst was becoming visible above the pointing horizon of the telescope. A total of 12 g'r'i'z' images with integration times of 66~s, 115~s and 375~s were obtained during the first night post burst. At the same time, NIR frames were acquired with a constant exposure of 10~s. In addition, the field was observed by GROND at days 1 and 2 after the trigger. Data reduction was performed using standard IRAF tasks \citep{tod93}, similar to the procedure outlined in \citet{kru08}. Absolute photometry of the afterglow was measured relative to field stars from the SDSS and 2MASS catalogs, which yields average accuracies of 0.04~mag in g'r'i'z' and 0.07~mag in JHK, and are reported in Tables \ref{table_data_R} and \ref{table_data}.

\begin{table*}
\centering
\caption{Table of optical/infrared data taken in other bands. GROND data in g', i', z' bands are transformed to Vega system. All magnitudes are corrected for local extinction due to the Milky Way, and errors are given at $1\sigma$.}
\label{table_data}
\begin{tabular}{cccccc|cccccc}
\hline
Mid time  & Exposure & Filter & Magnitude          &Telescope& \phantom{tex} & \phantom{tex} & Mid time  & Exposure & Filter & Magnitude        &Telescope\\
          & duration &        &                    &         & & &           & duration &        &                  &         \\
  (s)     & (s)      &        &                    &         & & &  (s)      & (s)      &        &                  &         \\
\hline
9042      & 66       &   g'   &   20.73 $\pm$ 0.05 & GROND   & & & 9266      & 66       &   z'   & 19.36 $\pm$ 0.05 &   GROND \\
9158      & 66       &   g'   &   20.75 $\pm$ 0.04 & GROND   & & & 9373      & 66       &   z'   & 19.44 $\pm$ 0.08 &   GROND \\
9266      & 66       &   g'   &   20.81 $\pm$ 0.06 & GROND   & & & 9514      & 115      &   z'   & 19.55 $\pm$ 0.10 &   GROND \\
9373      & 66       &   g'   &   20.80 $\pm$ 0.06 & GROND   & & & 9711      & 115      &   z'   & 19.58 $\pm$ 0.06 &   GROND \\
9514      & 115      &   g'   &   20.82 $\pm$ 0.05 & GROND   & & & 9903      & 115      &   z'   & 19.54 $\pm$ 0.05 &   GROND \\
9711      & 115      &   g'   &   20.86 $\pm$ 0.03 & GROND   & & & 10091     & 115      &   z'   & 19.53 $\pm$ 0.04 &   GROND \\
9903      & 115      &   g'   &   20.90 $\pm$ 0.03 & GROND   & & & 10423     & 375      &   z'   & 19.55 $\pm$ 0.04 &   GROND \\
10091     & 115      &   g'   &   20.90 $\pm$ 0.02 & GROND   & & & 10868     & 375      &   z'   & 19.59 $\pm$ 0.04 &   GROND \\
10423     & 375      &   g'   &   20.91 $\pm$ 0.02 & GROND   & & & 11320     & 375      &   z'   & 19.67 $\pm$ 0.02 &   GROND \\
10868     & 375      &   g'   &   20.95 $\pm$ 0.02 & GROND   & & & 11780     & 375      &   z'   & 19.66 $\pm$ 0.05 &   GROND \\
11320     & 375      &   g'   &   21.01 $\pm$ 0.02 & GROND   & & & 97805     & 1500     &   z'   & 22.28 $\pm$ 0.21 &   GROND \\
11780     & 375      &   g'   &   21.04 $\pm$ 0.02 & GROND   & & & 99600     & 1500     &   z'   & 22.51 $\pm$ 0.28 &   GROND \\
97805     & 1500     &   g'   &   23.42 $\pm$ 0.08 & GROND   & & & 184853    & 6000     &   z'   & 23.26 $\pm$ 0.25 &   GROND \\
99600     & 1500     &   g'   &   23.57 $\pm$ 0.10 & GROND   & & & 9214      & 240      &   J    & 18.69 $\pm$ 0.07 &   GROND \\
182130    & 1500     &   g'   &   24.18 $\pm$ 0.20 & GROND   & & & 9826      & 480      &   J    & 18.94 $\pm$ 0.05 &   GROND \\
183945    & 1500     &   g'   &   24.19 $\pm$ 0.15 & GROND   & & & 11124     & 1200     &   J    & 19.03 $\pm$ 0.04 &   GROND \\
185758    & 1500     &   g'   &   24.11 $\pm$ 0.13 & GROND   & & &  88.2     & 10.0     &   H    & 11.78 $\pm$ 0.05 &    REM  \\
187575    & 1500     &   g'   &   24.18 $\pm$ 0.13 & GROND   & & & 102.9     & 10.0     &   H    & 11.93 $\pm$ 0.03 &    REM  \\
9042      & 66       &   i'   &   19.78 $\pm$ 0.06 & GROND   & & & 117.6     & 10.0     &   H    & 12.38 $\pm$ 0.07 &    REM  \\
9158      & 66       &   i'   &   19.81 $\pm$ 0.05 & GROND   & & & 134.0     & 10.0     &   H    & 12.68 $\pm$ 0.04 &    REM  \\
9266      & 66       &   i'   &   19.74 $\pm$ 0.04 & GROND   & & & 152.1     & 10.0     &   H    & 12.67 $\pm$ 0.04 &    REM  \\
9373      & 66       &   i'   &   19.90 $\pm$ 0.06 & GROND   & & & 166.8     & 10.0     &   H    & 12.93 $\pm$ 0.03 &    REM  \\
9514      & 115      &   i'   &   19.92 $\pm$ 0.05 & GROND   & & & 182.4     & 10.0     &   H    & 13.05 $\pm$ 0.03 &    REM  \\
9711      & 115      &   i'   &   19.84 $\pm$ 0.04 & GROND   & & & 197.1     & 10.0     &   H    & 13.22 $\pm$ 0.07 &    REM  \\
9903      & 115      &   i'   &   19.92 $\pm$ 0.04 & GROND   & & & 212.6     & 10.0     &   H    & 13.35 $\pm$ 0.05 &    REM  \\
10091     & 115      &   i'   &   19.92 $\pm$ 0.05 & GROND   & & & 261.0     & 50.0     &   H    & 14.07 $\pm$ 0.07 &    REM  \\
10423     & 375      &   i'   &   20.01 $\pm$ 0.03 & GROND   & & & 340.5     & 50.0     &   H    & 14.32 $\pm$ 0.08 &    REM  \\
10868     & 375      &   i'   &   19.99 $\pm$ 0.02 & GROND   & & & 459.7     & 150.0    &   H    & 14.90 $\pm$ 0.06 &    REM  \\
11320     & 375      &   i'   &   20.06 $\pm$ 0.02 & GROND   & & & 9214      & 240      &   H    & $>$ 17.19        &   GROND \\
11780     & 375      &   i'   &   20.07 $\pm$ 0.02 & GROND   & & & 9826      & 480      &   H    & $>$ 17.79        &   GROND \\
97805     & 1500     &   i'   &   22.63 $\pm$ 0.12 & GROND   & & & 11124     & 1200     &   H    & 18.25 $\pm$ 0.07 &   GROND \\
99600     & 1500     &   i'   &   22.54 $\pm$ 0.12 & GROND   & & & 9214      & 240      &   K$_s$    & $>$ 16.48        &   GROND \\
184853    & 6000     &   i'   &   23.44 $\pm$ 0.15 & GROND   & & & 9826      & 480      &   K$_s$    & $>$ 17.18        &   GROND \\
9042      & 66       &   z'   &   19.37 $\pm$ 0.08 & GROND   & & & 11124     & 1200     &   K$_s$    & 17.59 $\pm$ 0.07 &   GROND \\
9158      & 66       &   z'   &   19.44 $\pm$ 0.07 & GROND   & & &           &          &        &                  \\
\hline
\end{tabular}
\end{table*}

\subsubsection{Temporal and spectral behaviors}

The optical light curve can be fit using a broken power law model, with a break time on the order of 1000 s. As can be seen in Fig. \ref{lc_all}, the light curve flattens at that time. It is worth noting that this flattening cannot be due to the host galaxy, fainter than R$\sim24$ \citep{lev09} when the afterglow magnitude is of the order of R$\sim19-20$. Before the break, the decay index is $\alpha_{1,O} = 1.50 \pm 0.06$, { steeper} than the X-ray decay index at the same date (see Table \ref{table_061126}). After the break, the decay index is $\alpha_{2,0} = 0.97 \pm 0.03$, flatter than the X-ray decay index at the same time. We do not observe any color variation (both in optical and in X-rays) before and after the temporal break seen in optical (see Table \ref{table_color}). If we assume the host extinction to be null, the value of R $-$ H = 2.4 implies $\beta_{O, obs} = 1.32$. A non zero extinction would cause a flatter $\beta_{O}$, thus one should consider this as an upper limit: $\beta_{O} < 1.32$. The value of the optical-to-X-ray spectral index is $\beta_{OX} > 0.53 \pm 0.04$. 

\begin{table}
\caption{\label{table_061126} Summary of the properties of GRB 090102.} 
\centering                          
\begin{tabular}{cc}        
\hline                 
Property                        & GRB 090102 \\
\hline
Early optical decay index            & $1.50 \pm 0.06$\\
Late optical decay index             & $0.97 \pm 0.03$\\
X-ray decay index                    & $1.34 \pm 0.02$\\
X-ray spectral index            & $0.83 \pm 0.09$\\
\hline
\end{tabular}
\end{table}

\begin{table}
\centering
\caption{GRB 090102 afterglow R$-$H color. All magnitudes are corrected for local extinction due to the Milky Way.}
\label{table_color}
\begin{tabular}{cc}
\hline
Mid time  & R $-$ H \\
  (s)     & color \\
\hline
 88.2 &  2.42 $\pm$ 0.25 \\
102.9 &  2.47 $\pm$ 0.33 \\
152.1 &  2.53 $\pm$ 0.24 \\
166.8 &  2.25 $\pm$ 0.23 \\
197.1 &  2.48 $\pm$ 0.27 \\
11124 &  2.39 $\pm$ 0.08 \\
\hline
\end{tabular}
\end{table}

\begin{figure}
\centering
  \includegraphics[width=7.5cm]{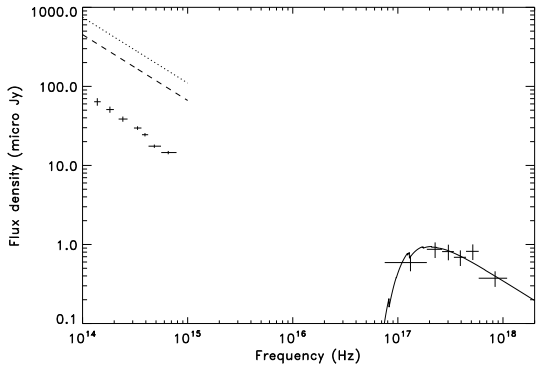}
  \caption{\label{fig_sed}Spectral energy distribution of the afterglow of GRB 090102 observed 11.3 ks after the burst. Frequencies are expressed in the observer frame. The solid line represent the best fit X-ray model. Dotted line is the direct extrapolation of this model toward optical frequencies (assuming no extragalactic extinction). As there may be an X-ray flare at that time, we also rescaled this extrapolation (dashed line) to the level of the X-ray continuum at the same time.}
\end{figure}

\subsubsection{Optical extinction}

\begin{table}
\centering
\caption{ GRB 090102 optical SED fit result. We indicate for various models the best fit parameters and the goodness of the fit. Within the fit, the $\beta_O$ parameter was not allowed to become negative. In all cases, the redshift was fixed to 1.547.}
\label{table_sed}
\begin{tabular}{ccccc}
\hline
Model                 & $A_v^{\rm host}$            & $\beta_O$ & $\chi^2_\nu$ & d.o.f. \\
\hline
Milky Way             &  0.35 $\pm$ 0.29        & 0.52$^{+0.32}_{-0.36}$ & 0.07 & 4\\
Large Magelanic Cloud &  $0.52^{+0.29}_{-0.38}$ & 0.29$^{+0.50}_{-0.28}$ & 0.07 & 4\\
Small Magelanic Cloud &  $0.04^{+0.13}_{-0.04}$ & 0.93$^{+0.13}_{-0.22}$ & 1.35 & 4\\
Calzetti law          &  $0.77^{+0.23}_{-0.77}$ & 0.1$^{+1.0}_{-0.1}$    & 1.19 & 4\\
\hline
\end{tabular}
\end{table}

\begin{figure}
\centering
  \includegraphics[width=7.5cm, angle=-90]{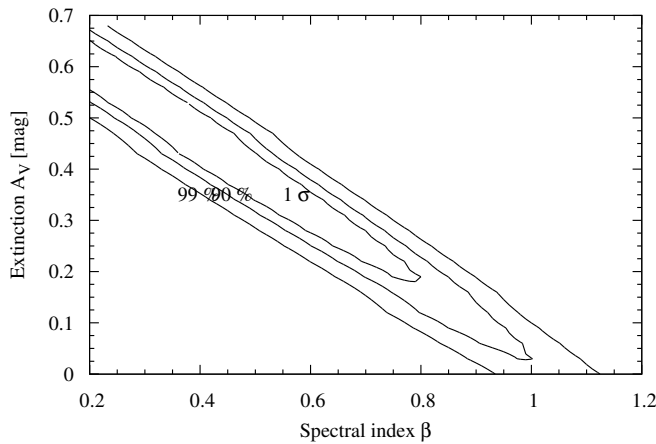}
  \caption{\label{fig_referee} Contour plot in the $\beta_O$-$A_v$ plane for a Large Magellanic Cloud extinction model. Contours are indicated for the 68\%, 90\% and 99\% confidence levels. As one can see from the pronounced "banana-shape" of the plot, there is a strong degeneracy between these two parameters.}
\end{figure}

The estimation of the optical extinction is a non-trivial problem, as we need to assume a spectral slope of the afterglow continuum in order to deduce the amount of extinction needed in the optical to reproduce that continuum \citep[see e.g.][]{str09}. In this article we prefer to focus on the constraints we can set on the GRB models, and thus we cannot make any assumption on the afterglow continuum. As a matter of consequence, we have tried to fit the optical-near infrared SED without using the X-ray information, assuming a simple power-law model\footnote{One should think that this is already an hypothesis on the data, i.e. that there is no spectral break located between the g' and the K bands.}. We have tested the most common models used, i.e. the Milky Way, the Large and Small Magellanic Clouds and the Calzetti law \citep[see][and references therein]{cal94, str04}. Results are reported in Table \ref{table_sed}. As can be seen from that Table, all models are acceptable. However, the small value of $\chi^2_\nu$ reported in some case indicates a statistical problem. In fact, the spectral index and the host extinction are degenerate parameters, as can be seen from Fig. \ref{fig_referee}: if the errors are defined in Table \ref{table_sed}, this is only due to the fact that both $A_v$ (for physical reason) and $\beta_O$ (by hypothesis) are not allowed to become negative. As a matter of consequence, we cannot derive the host extinction from the optical photometric data alone.

However, we are able to provide two comments. 
First, in the host frame, we have an X-ray measurement of $N_H = 8.8_{-2.5}^{+2.7} \times 10^{21}$ cm$^{-2}$. This may imply that significant local extinction is present. Within our galaxy, this would lead to $A_{\rm V} = 4.9_{-1.4}^{+1.5}$, that can be considered as an upper limit \citep[see the discussion of][]{str04}.

Finally, in the host frame we have measurements ranging from 180.1 nm to 852.2 nm, i.e. covering the UV to $\sim$ I bands. This last band should be the least affected by the host extinction. This is, however, not a band free of extinction, and cannot be used to derive an estimation of the continuum flux density. Assuming that the optical and X-ray emissions are produced by the same mechanism, the true value is located between the data points and the dotted line on Fig. \ref{fig_sed}, corresponding to an extinction range of 0 -- $\sim 2.4$ mag.

\section{Model fitting}
\label{sec_model}
\subsection{Standard model}

\begin{table*}
\caption{\label{table_closure} Closure relationships in the standard fireball model computed using the spectral and temporal information from the X-ray band. The specific frequency is $\nu_c$ (the cooling frequency) in the slow cooling regime, or $\nu_m$ (the injection frequency) in the case of fast cooling. All errors are quoted at the 90 \% confidence level. Due to the absence of measured spectral index in the optical, optical values of the relationships are not reported} 
\centering                          
\begin{tabular}{ccccccc}        
\hline                 
Medium class    & Cooling & Specific    & Closure       &  \multicolumn{3}{c}{Observed value} \\
and             & regime  & frequency   & relationship  &  \multicolumn{3}{c}{X-rays (0.5-10.0 keV)} \\
geometry        &         & position    &                               & (single power  & (early broken  & (late broken\\
                &         &             &                               & law hypothesis)& power law)     & power law) \\
\hline
Isotropic Wind  & Fast    & $\nu_m<\nu_X$ & $\alpha - 1.5 \beta  + 0.5$ = 0 & $0.6\pm0.2$ & $0.5\pm0.2$ & $0.7\pm0.3$\\
                &         & $\nu_m>\nu_X$ & $\alpha - 0.5 \beta$        = 0 & $0.9\pm0.1$ & $0.9\pm0.1$ & $1.1\pm0.2$\\
                & Slow    & $\nu_c<\nu_X$ & $\alpha - 1.5 \beta + 0.5$  = 0 & $0.6\pm0.2$ & $0.5\pm0.2$ & $0.7\pm0.3$\\
                &         & $\nu_c>\nu_X$ & $\alpha - 1.5 \beta  - 0.5$ = 0 &$-0.4\pm0.2$ &$-0.5\pm0.2$ & $-0.3\pm0.3$\\
Isotropic ISM   & Fast    & $\nu_m<\nu_X$ & $\alpha - 1.5 \beta + 0.5$  = 0 & $0.6\pm0.2$ & $0.5\pm0.2$ & $0.7\pm0.3$\\
                &         & $\nu_m>\nu_X$ & $\alpha - 0.5 \beta$        = 0 & $0.9\pm0.1$ & $0.9\pm0.1$ & $1.1\pm0.2$\\
                & Slow    & $\nu_c<\nu_X$ & $\alpha - 1.5 \beta + 0.5$  = 0 & $0.6\pm0.2$ & $0.5\pm0.2$ & $0.7\pm0.3$\\
                &         & $\nu_c>\nu_X$ & $\alpha - 1.5 \beta$        = 0 & $0.1\pm0.2$ & $0.0\pm0.2$ & $0.2\pm0.3$\\
Jetted fireball & Slow    & $\nu_c<\nu_X$ & $\alpha - 2 \beta$          = 0 &$-0.3\pm0.2$ &$-0.4\pm0.3$ &$-0.2\pm0.3$\\
                &         & $\nu_c>\nu_X$ & $\alpha - 2 \beta - 1.0$    = 0 &$-1.3\pm0.2$ &$-1.4\pm0.3$ &$-1.2\pm0.3$\\
\hline
\end{tabular}
\end{table*}

We first used the prescriptions of \citet{pan00} and the X-ray data alone in order to test the standard model. They are summarized in Table \ref{table_closure} \citep[from][]{sar98, sar99, rho97}. As one can see from this Table, in order to accommodate the standard model and the measurements made in X-ray, we should have the cooling frequency located { above} the X-ray band during the entirety of the observation. In such a case, a value of $p = 2\beta_X = 2.7 \pm 0.2$ can fit both the spectral and temporal X-ray indices if the surrounding medium has a constant density. However, a constant density medium implies that the cooling frequency is decreasing as $t^{-0.5}$. Any crossing of this frequency within the X-ray band would appear as a simultaneous increase in the spectral index by 0.5 and in the temporal decay index by 0.25. While the latter could be accommodated by the broken power law model, the former is not supported by the data. This implies that the cooling frequency should be above the X-ray band up to $\sim 5 \times 10^5$ s after the event, or above { 800 keV} at the start of the X-ray observation \citep[thus implying very small values of $\epsilon_B$ and/or $E_{\rm tot}$, the energy fraction located within the magnetic field and the total energy of the fireball respectively][]{pan00}. This is unlikely; moreover, in such a case, according to the model, the optical light curve should decay either with the same slope than the X-ray light curve or with a decay index of 0.5 \citep{pan00}: both hypotheses are ruled out by the data.

In conclusion, the standard model alone, in its simplest expression, cannot reproduce the global spectral and temporal behavior of the afterglow of GRB 090102.

\subsection{Two component models}

The afterglow of GRB 061126 features the same temporal and spectral behavior as the one of GRB 090102 \citep[][see also Table \ref{table_061126}]{per08, gom08}. These authors used a multi-component model rather than the standard fireball model in order to explain the broad-band data of this burst. We thus checked if inserting an additional component to the fireball model could reproduce the data of the afterglow of GRB 090102.

The first step was to decide in which band we see the signature of the extra-component, and in which band the fireball model alone can reproduce both the spectral and temporal properties. In the standard fireball model, the fact that the optical and X-ray light curves do not present the same decay implies that a specific frequency \citep[either the cooling or the injection frequency, see their definition in][]{mes06} is located between the optical and X-ray bands. Below the injection frequency, the decay index should be 0.5. With an optical decay of $\alpha_{1,O} = 1.50 \pm 0.06$, we can rule out this hypothesis, and deduce that the cooling frequency needs to be located between the optical and X-ray bands. This is not compatible with the X-ray data (see Table \ref{table_closure}): an extra component in the X-ray band is the only solution to solve this issue. 

Under the above assumption, the optical band can be described by the fireball model and it is the X-ray band that needs an extra component in addition to the fireball in order to be described. This hypothesis needs to be confirmed by checking if the optical band can indeed be fit with the standard model. In fact, the flattening of the optical light curve around 1 ks is not possible in the standard model. However, several "common additions" to the standard model can provide an explanation: the addition of a reverse shock in the early optical data or a change in the surrounding medium density profile \citep[the termination shock, see][]{che04,ram01}. In the latter case, we can derive a value of $p = 4/3 \alpha_{2,O} + 1 = 2.29 \pm 0.04$. The physical position of the termination shock cannot be constrained due to the presence of a second component in the X-ray data. We can thus conclude that our hypothesis is valid: the optical band can be fit with the standard model.

Lastly, one then needs to provide for the physical nature of the additional X-ray component. The requirements are that it needs to be:  non-thermal, at all times brighter than the forward shock emission responsible for the optical emission, decaying monotonously without spectral variation, and producing few (if any) optical emissions. We do not conclude on the mechanism that produces these extra photons in X-ray. However, the latter requirement is the more constraining one, as already noted by \citet{per08} for GRB 061126: it is quite difficult to accommodate a mechanism that produces X-rays without also producing optical photons. Note additionally that late in the light curve (after 1 ks), the standard model implies that the X-ray spectral index is $p/2 \sim 1.2$. The additional component needs to produce hard photons in order to obtain $\beta_X = 0.83 \pm 0.09$. 

\subsection{The cannonball model}

In the cannonball model, GRBs and their afterglows are produced by bipolar jets of highly relativistic plasmoids (CB) of ordinary matter ejected in the birth of neutron stars or black holes \citep[see e.g.][and references therein]{dad09}. 

A prompt, hard X-ray/gamma-ray pulse is produced by the thermal electrons in the CB plasma, via inverse Compton scattering of photons emitted/scattered into a cavity created by the wind/ejecta blown from the progenitor star or a companion star long before the GRB itself. Slightly later, when the CB encounters the wind/ejecta, the electrons of the ionized gas in front of it are swept into the CB and are Fermi accelerated by its turbulent magnetic field, emitting synchrotron radiation that dominates the prompt optical emission and the broad band afterglow emission. Thus, the observed emissions are due to two components: a short lasting Inverse Compton Scattering component, seen only at high energy, and a classical synchrotron emission \citep{dad09b}. 

Following the notations and prescriptions of \citet{dad09}, we can explain all the X-ray data assuming that we observe the synchrotron emission of a CB traveling within a medium of constant density (the so-called ISM) with the bend frequency located below the X-ray band. From the model, which gives, in that case $\beta_X = p/2$, we can derive $p = 1.7 \pm 0.2$. This value of $p$ implies $\alpha_X = \beta + 0.5 = p/2 + 0.5 = 1.35 \pm 0.1$, in perfect agreement with our finding ($\alpha_X = 1.34 \pm 0.02$). We note that this value of $p$ is lower than 2, and thus there should be a break/cutoff located at high energy so that the electron energy remains finite. 

In the optical, things are more complicated. According to \citet{dad09}, the spectral index in the optical should evolve from $\beta_O  \approx 0.5$ during the prompt emission to  $\beta_O \approx 0.83 \pm 0.09$ (i.e. the same as that  in X-ray) during the late-time afterglow. Both values are compatible with the constraint derived from the color of the afterglow ($\beta \leq 1.32$). During the prompt optical emission, the decay index should be $\alpha_O = \beta_O + 1 \approx 1.5$ \citep[see Eq.~33 in][]{dad09}. This is consistent with the observed value of $\alpha_{1,O} = 1.50 \pm 0.06$. The flattening of the optical light curve may be explained by a transition from a wind environment to an ISM (as in the fireball model, see above). In such a case, the light curve should reach an asymptotic decay with $\alpha_{2,O} = \alpha_X = 1.34 \pm 0.09$ \citep[see Eqs.~29 and 30 in][]{dad09}. We observed a value of $\alpha_{2,O} = 0.97 \pm 0.03$. However, the value of $\alpha_{2,O} = 1.34 \pm 0.09$ is only an asymptotic value, and the observed could vary if the observation is close to the time of the crossing of the bend frequency. In fact, we should observe an effective temporal decay index $\alpha_2(t) = \beta_O(t)+1/2$ evolving from $\alpha_2 \sim 1$ to $\alpha_2 = 1.34 \pm 0.02$. Being close to the asymptotic value, the measurement is marginally compatible with the CB model expectation.  Lack of late time temporal and spectral data prevents a confirmation of this prediction. Indeed, once reaching the asymptotic values, the CB model indicates that we should observe the same spectral and temporal behavior for the optical and X-ray bands. Looking to Fig. \ref{fig_sed}, this is hardly supported by the data. However, a large gray extinction, with A$_I \sim 2.5$ would make compatible the optical and X-ray emissions. In the absence of information about the host extinction, we cannot rule out such an extinction, and the CB model may indeed reproduce all the data with the expense of a few fine tunings.

\section{Discussion and Conclusions}
\label{sec_discu}

We have presented observations of the afterglow of GRB~090102. We used data taken by optical telescopes and by the XRT on board the \emph{Swift} spacecraft. The light curve of this event features an unusual behavior. In X-rays, it presents a very smooth light curve with no hint of a temporal break. In the optical, the light curve presents a steep-flat behavior, with a break time at $\sim$ 1 ks after the burst. This unusual light curve was also observed with GRB 061126. The standard fireball model cannot reproduce the observations without the addition of several components. The addition of a component at high energy solves the discrepancies between the optical and X-ray emissions. The optical data could then be interpreted as either due to a termination shock, locating the end of the free-wind bubble at the position of the optical break; or as a normal fireball expanding in an ISM, with a reverse shock present at an early time (less than 1 ks after the event). Any radio observations that could constrain the ISM density value would also constrain the wind density, and thus a key property of the progenitor of this GRB. However, the weak point of this model is that requires strong fine tuning in order to suppress the optical emission of the additional component, responsible for the X-ray emission. The cannonball model can also partly reproduce the data. It appears clear, however, that in order to explain the broad-band emission, some fine-tuning of this model is mandatory, likewise for the fireball model.

The strange behavior of GRB 090102 provides a useful test-event for models that attempt to reproduce and explain the GRB phenomenon.  The choice of the models tested here was strongly limited by the very small number of models that have published formulae of the temporal and spectral indices for GRB afterglows. Of course, we do not pretend to be exhaustive in our testing as the array of alternative models is huge. {\it Swift} has shown the diversity of GRB afterglows, and it may now be time for observers to test and reduce the diversity of GRB afterglow models.

\section*{Acknowledgments}

The authors thank G. Stratta for help during X-ray data analysis, and A. Dar for his kind and valuable help during the cannonball model analysis. Thanks are also due to K. Dzurella for help during the language improvement of the paper. B. Gendre acknowledges support from {\it Centre National d'Etudes Spatiales} (CNES). TK acknowledges support by the DFG cluster of excellence 'Origin and Structure of the Universe'.

\label{lastpage}

\end{document}